\documentclass[letterpaper, 10 pt, conference]{ieeeconf}  

\IEEEoverridecommandlockouts                              

\usepackage{graphicx} 
\usepackage{amsmath} 
\usepackage{amssymb}  
\usepackage{physics}
\usepackage{xcolor}
\usepackage{soul}
\usepackage{multirow}
\usepackage[normalem]{ulem} 

\DeclareMathOperator*{\minimize}{minimize}

\usepackage{longtable,tabularx} 

\title{\LARGE \bf
Fast Assignment in Asset-Guarding Engagements using Function Approximation
}

\author{Neelay Junnarkar$^{1}$, Emmanuel Sin$^{2}$, Peter Seiler$^{3}$, Douglas Philbrick\(^{4}\), and Murat Arcak$^{1}$
\thanks{*This work was supported by Office of Naval Research under grant N00014-18-1-2209.}
\thanks{$^{1}$Neelay Junnarkar and Murat Arcak are with the Department of Electrical Engineering and Computer Sciences, University of California at Berkeley, Berkeley, CA 94720
USA (e-mail: neelay.junnarkar@berkeley.edu; arcak@berkeley.edu).}
\thanks{$^{2}$ Emmanuel Sin is with the Department of Mechanical Engineering,
University of California at Berkeley, Berkeley, CA 94010 USA (e-mail:
emansin@berkeley.edu).}
\thanks{$^{3}$Peter Seiler is with the Department of Electrical Engineering and
Computer Science, University of Michigan at Ann Arbor, Ann Arbor,
MI 48109 USA (e-mail: pseiler@umich.edu).}%
\thanks{\(^{4}\)Douglas Philbrick is with the U.S. Naval Air Warfare Center-
Weapons Division (e-mail: douglas.philbrick@navy.mil).}
}

\begin{document}

\maketitle
\thispagestyle{empty}
\pagestyle{empty}

\begin{abstract}


This letter considers assignment problems consisting of \(n\) pursuers attempting to intercept \(n\) targets. We consider stationary targets as well as targets maneuvering toward an asset. The assignment algorithm relies on an \(n\times n\) cost matrix where entry \((i,j)\) is the minimum time for pursuer \(i\) to intercept target \(j\). Each entry of this matrix requires the solution of a nonlinear optimal control problem. This subproblem is computationally intensive and hence the computational cost of the assignment is dominated by the construction of the cost matrix. We propose to use neural networks for function approximation of the minimum time until intercept. The neural networks are trained offline, thus allowing for real-time online construction of cost matrices. Moreover, the function approximators have sufficient accuracy to obtain reasonable solutions to the assignment problem. In most cases, the approximators achieve assignments with optimal worst case intercept time. The proposed approach is demonstrated on several examples with increasing numbers of pursuers and targets.
\end{abstract}

\section{INTRODUCTION}

Missile allocation problems arise in defense scenarios where a group of missiles must be assigned to targets. An overview of various missile allocation problems is given in \cite{matlin-map} and \cite{isaacs}. A typical type of problem is where a group of missiles must defend an asset from a group of incoming threats before any of the threats reach the asset. In \cite{sin-opt-assign} this is extended to a group of \(n\) collaborative pursuer missiles defending a single asset from a group of \(m \leq n\) evading threat missiles, and presents an assignment problem formulation to match pursuers to evaders.

\begin{figure}[t]
\centering
\includegraphics[width=.45\textwidth]{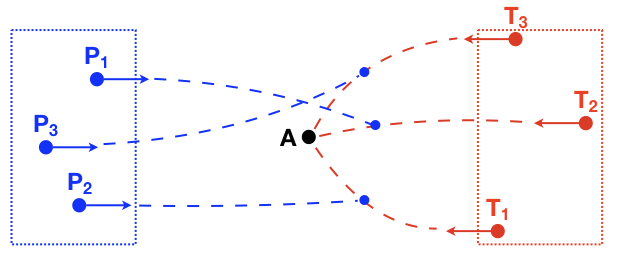}
\caption{\(3 \times 3\) Engagement with Maneuvering Targets. The dashed boxes represent the regions of possible initial conditions, with blue for the pursuers and red for targets. The blue dots labeled \(P_i\) and the red dots labeled \(T_i\) are the three pursuers and targets respectively at their starting locations. The arrows represent the initial velocities of the pursuers and targets. The red dashed lines are the targets' trajectories toward the asset, \(A\), and the blue dashed lines are the pursuers' intercept trajectories, where the intercept points are the small blue dots. }
\label{fig:engagement}
\end{figure}

In this letter we consider engagements with \(n\) pursuers and \(n\) targets. Figure \ref{fig:engagement} depicts a \(3 \times 3\) engagement with maneuvering targets where all three targets are intercepted by the pursuers. Within a single engagement, we allow either all targets to be stationary, or all targets to be maneuvering toward an asset. Maneuvering targets follow proportional navigation guidance, a widely used guidance law, toward the asset \cite{Zarchan1990TacticalAS}. The problem is to assign pursuers to targets such that pursuers intercept the targets in minimum overall time. 

An existing method to solve such missile allocation problems is to use a framework where a cost is associated with each of the \(n^2\) possible pairs of pursuers and targets, and then an assignment problem is solved to determine which particular assignment of pursuers to targets is used \cite{sin-opt-assign} \cite{bakolas2021} \cite{weintraub2020}. An overview of various task assignment problems, in which agents are allocated to tasks, is given in \cite{pentico}. Reference \cite{burkard_ch4} gives an overview of methods to solve the linear sum assignment problem, where agents are assigned such that the sum of assigned costs is minimized. Reference \cite{burkard_ch6} gives an overview of solution methods to the bottleneck assignment problem, where agents are assigned such that the maximum assigned cost, the bottleneck, is minimized. While papers such as \cite{nazemi2017}, \cite{liu2009}, and \cite{chen2008} develop methods to accelerate solving large assignment problems given pre-computed cost matrices, the time required to compute the cost matrix itself remains a computational obstacle. Computing practically useful rewards often requires solving computationally expensive optimal control problems \cite{sin-opt-assign} \cite{bakolas2021}, which makes these methods impractical for real-time applications.

Our main contribution is in the acceleration of a suitable cost matrix with which to solve assignment problems for allocating pursuers to stationary or maneuvering targets. We formulate an assignment problem that assigns pursuers to targets to minimize the maximum assigned intercept time. This involves predicting the minimum intercept time for each pair of pursuer and target. We present an optimal control method to compute this intercept time. To accelerate this method, we train a neural network offline to approximate the output of the optimization program. This enables rapid online computation of the cost matrix used in the assignment problems.

Neural networks have been used in accelerating computationally expensive operations by acting as function approximators, such as in \cite{zhu2019} where they are used to quickly approximate the optimal minimum fuel cost of orbital transfers for interplanetary missions.  Our approximation of the minimum intercept time proceeds in two steps: first, a model which predicts whether a particular pursuer can intercept a particular target; and second, a model which predicts the minimum time until intercept for a pursuer and target pair, assuming that interception is possible. We demonstrate the accuracy of this approach by comparing the resulting assignment bottleneck values to the bottleneck values of the assignments computed using the true minimum intercept times.

The methods presented here are of broader interest to scenarios that require real-time combinatorial optimization, where large numbers of values must be estimated quickly. Such scenarios might include car sharing programs where cars must be assigned to users, wildfire fighting where one must decide which front to fight to maximize containment efforts, and evacuations where teams must be dispatched to numerous locations. In these scenarios, the highly dynamic environment may make optimal control or combinatorial optimization techniques unviable.

\section{ASSET GUARDING ENGAGEMENTS} \label{sec:asset-guarding-engagements}

In this section we introduce the two types of engagements addressed in this paper. We then formulate a trajectory optimization problem and an assignment problem to determine optimal assignments for pursuers to intercept targets.

\subsection{Engagement Type}

We consider two types of engagements: 
\begin{enumerate}
    \item assigning \(n\) pursuers to \(n\) stationary targets
    \item assigning \(n\) pursuers to \(n\) maneuvering targets guided by proportional navigation (PN) towards a single stationary asset.
\end{enumerate}
In both engagement types, we seek to minimize the engagement time, i.e., the time required for all targets to be intercepted, by optimally assigning pursuers to targets. Each pursuer is assumed to take a minimum-time trajectory towards its assigned target. We also assume that each pursuer can only intercept a single target and that the pursuers act independently of each other. Hence, the assigned pursuer-target pair that requires the longest intercept time decides the overall engagement time.

We solve these problems by first computing all \(n^2\) pursuer-target pairwise minimum intercept times, and then solving a bottleneck assignment problem to minimize the maximum assigned time. Section \ref{sec:traj-opt} describes an optimal control problem to compute the minimum intercept time for a single pursuer-target pair. Section \ref{sec:assignment-prob} describes the bottleneck assignment problem which takes as input a matrix of minimum intercept times, as computed by the optimal control problem, and outputs an assignment which minimizes the maximum assigned time. 

\subsection{Trajectory Optimization Problem Formulation} \label{sec:traj-opt}

Given a pursuer-target pair \((P_i, T_j)\), an optimal control problem is solved to determine the trajectory that a pursuer should take to intercept a target in minimum time. The problem takes in the following inputs:
\begin{itemize}
    \item initial condition of a pursuer, $\mathbf{r}^{P_i}_0$, $\mathbf{v}^{P_i}_0$
    \item position trajectory of a target, $\bar{\mathbf{r}}^{T_j}$.
\end{itemize}
In the case of a maneuvering target, its position trajectory is pre-computed in simulation given its initial condition and the assumption that it uses a PN guidance law towards the asset. Since the asset is non-maneuvering, moreover stationary, we assume that the target uses a navigation ratio value of 3, which has been shown to be energy-optimal for linear engagement kinematics models \cite{PNratio}.

The problem formulation is as follows:
\begin{align}
J^*_{TrajOpt}( \mathbf{r}^{P_i}_0, \mathbf{v}^{P_i}_0, \bar{\mathbf{r}}^{T_j} ) \ = \
\minimize_{ \mathbf{r}, \mathbf{v}, \mathbf{u}, t_f } \hspace{0.2cm} \int_{0}^{t_f} 1 \ dt \hspace{0.5cm} \label{eqn:OCP_objective}
\end{align}
\begin{align}
\text{ s.t.} \ \ &\dot{\mathbf{r}}(t) = \mathbf{v}(t)  
											&\forall \ t \in [0, t_f] \label{eqn:OCP_rdyn}\\
&\dot{\mathbf{v}}(t) = \mathbf{u}(t) - \beta \rho(\mathbf{r}(t)) \lVert \mathbf{v}(t) \rVert \mathbf{v}(t) 
											&\forall \ t \in [0, t_f] \label{eqn:OCP_vdyn} \\
&\lVert \mathbf{u}(t) \rVert_{\infty} \leq u_{max}													&\forall \ t \in [0, t_f]  \label{eqn:OCP_umax} \\
&r_z(t) \geq 0		
											&\forall \ t \in [0, t_f]  \label{eqn:OCP_height} \\
&\mathbf{r}(0) = \mathbf{r}^{P_i}_0	 \label{eqn:OCP_ric} \\
&\mathbf{v}(0) = \mathbf{v}^{P_i}_0	\label{eqn:OCP_vic} \\		
&\lVert \mathbf{r}(t_f) - \bar{\mathbf{r}}^{T_j}(t_f) \rVert \leq r_{capture}. \label{eqn:OCP_capture} \end{align}

We model the pursuer (and target) as a 3-degree-of-freedom point mass whose motion is affected by control accelerations and atmospheric drag. Using the Cartesian coordinate system, position is measured with respect to an inertial reference frame placed on a ``flat Earth'' surface such that the positive z-direction is pointing ``up." In our dynamical model of the pursuer vehicle, the state vector composed of $\mathbf{r} := [r_x \ r_y \ r_z]^\top$ and $\mathbf{v} := [v_x \ v_y \ v_z]^\top$ represents the pursuer's position [ft] and velocity [ft/s], respectively. The input to the system $\mathbf{u} := [u_x \ u_y \ u_z]^\top$ represents the acceleration [ft/s$^2$] that may be applied to the pursuer vehicle, through thrust or actuated aerodynamic surfaces. We assume that each channel of the acceleration is bounded in magnitude (i.e., $\lVert \mathbf{u}(t) \rVert_{\infty} \leq u_{max}$).
Atmospheric density [slug/ft$^3$] is an exponential function of the altitude: $\rho(\mathbf{r}) := \rho_0 \exp(- r_z / h_0)$, where $\rho_0$, the atmospheric density at reference altitude $h_0$, is given. We assume a constant mass $m$  [slug], drag coefficient $C_D$, reference area $S$ [ft$^2$], and maximum acceleration $u_{max}$ [ft/s$^2$].  We define the ballistic coefficient as $\beta := \frac{C_{\scriptscriptstyle D}S}{2m}$. 

The pursuer's objective is to reach within a Euclidean distance $r_{capture}$ [m] of the target's position in minimum time given its initial condition, dynamical constraints, input constraints, and state constraints (i.e., maintain positive altitude). We solve the trajectory optimization problem described by equations (\ref{eqn:OCP_objective}) - (\ref{eqn:OCP_capture}) using sequential convex programming (SCP). Recent advances in SCP have enabled efficient computation of locally optimal trajectories for nonlinear systems with non-convex constraints and objectives \cite{SCP}. SCP is an iterative method that repeatedly formulates and solves a convex, finite-dimensional parameter optimization problem that approximates the original non-convex optimal control problem. Convex formulations are typically achieved by linearizing the nonlinear system dynamics and constraints about a nominal trajectory (e.g., the solution from a previous iteration). Fast and reliable Interior Point Method algorithms \cite{Nocedal} may be used to solve these convex subproblems.

A solution to the trajectory optimization problem provides the pursuer's minimum-time intercept trajectory $\{ \bar{\mathbf{r}}^*(t), \bar{\mathbf{v}}^*(t), \bar{\mathbf{u}}^*(t) \}, \ t \in [0, \bar{t}^*_f]$. If the pursuer's minimum intercept time is greater than the time required for the target to reach the asset, or if the pursuer is unable to intercept the target, then we consider this pursuer and target pair as infeasible in terms of interception. 

\subsection{Assignment problem formulation} \label{sec:assignment-prob}

In this section, we formulate an assignment problem to assign \(n\) pursuers to \(n\) targets. The problem takes as input an \(n \times n\) cost matrix \(C\) where the element \(C_{i,j}\) is the minimum time required for pursuer \(P_i\) to intercept target \(T_j\):
\begin{equation}
C_{i,j} := J^*_{TrajOpt}( \mathbf{r}^{P_i}_0, \mathbf{v}^{P_i}_0, \bar{\mathbf{r}}^{T_j} ).
\end{equation}
The assignment problem has a \(n \times n\) matrix \(Z\) as the decision variable, where the element \(Z_{i,j}\) takes on a value 1 if pursuer \(P_i\) is assigned to target \(T_j\) and 0 otherwise.
Each pursuer is assigned to exactly one target.

We define the cost associated with a given assignment as the maximum of the assigned intercept times:
\begin{align*}
    J_{\text{bottleneck}}(C, Z) = \max_{i,j} \{C_{i,j}Z_{i,j}\}.
\end{align*}
This is the bottleneck cost of the assignment \(Z\) on cost matrix \(C\). If pursuer \(P_{i}\) and target \(T_{j}\) are such that \(Z_{i,j} = 1\) and \(J_{\text{bottleneck}}( C, Z ) = C_{i,j}\), then the pursuer-target pair \((P_i, T_j)\) is considered a bottleneck of the assignment problem with bottleneck time \(C_{i,j}\).

The bottleneck assignment problem (BAP) is to find an assignment \(Z\) which minimizes the maximum assigned intercept time \(J_{\text{bottleneck}}(C, Z)\) given the cost matrix \(C\). Let \(Z_{BAP}^*(C)\) be such an assignment. The bottleneck assignment problem is given formally by the following optimization problem:
\begin{align}
&\minimize_{ Z } \ \ \max_{i,j} \{ C_{\scriptscriptstyle i,j} Z_{\scriptscriptstyle i,j} \} \label{eqn:assign-obj} \hspace{1.5cm} \\
&\text{ s.t.} \ \ \sum_{i=1}^{N} \hspace{0.1cm} \ Z_{\scriptscriptstyle i,j} = 1  \hspace{0.55cm} \forall \ j = 1, \ldots, N  \label{eqn:assign-target-one}\\
&\phantom{\text{s.t. }} \ \ \sum_{j=1}^{N} \hspace{0.1cm} \ Z_{\scriptscriptstyle i,j} = 1 \hspace{0.55cm} \forall \ i = 1, \ldots, N  \label{eqn:assign-pursuer-one}\\
&\phantom{\text{s.t. }} \hspace{0.95cm} \ Z_{\scriptscriptstyle i,j} \in \{ 0, 1 \}. \label{eqn:assign-z-binary}
\end{align}
The constraints in equations \eqref{eqn:assign-target-one} and \eqref{eqn:assign-pursuer-one} ensure each target has exactly one pursuer assigned to it and that each pursuer is assigned to exactly one target, respectively. 

\section{Minimum Time Until Intercept Approximator} \label{sec:approximator}

The computationally limiting factor of solving for an assignment as presented in Section \ref{sec:asset-guarding-engagements} is in computing the  \(n^2\) minimum intercept times \(J^*_{TrajOpt}( \mathbf{r}^{P_i}_0, \mathbf{v}^{P_i}_0, \bar{\mathbf{r}}^{T_j} )\) needed to construct the cost matrix \(C\). In this section we present a neural network based model to approximate \(J^*_{TrajOpt}( \mathbf{r}^{P_i}_0, \mathbf{v}^{P_i}_0, \bar{\mathbf{r}}^{T_j} )\). We describe the structure of the function approximator, how the neural networks are trained offline, and how the function approximator is used to produce assignments.

\subsection{Function Approximator Structure}

The function approximator, \(\tilde{f}\), takes as input a pursuer and target's initial conditions:
\begin{align*}
    \tilde{f}( \mathbf{r}^{P_i}_0, \mathbf{v}^{P_i}_0, \bar{\mathbf{r}}^{T_j}_0,  \dot{\bar{\mathbf{r}}}^{T_j}_0) \approx 
        J^*_{TrajOpt}( \mathbf{r}^{P_i}_0, \mathbf{v}^{P_i}_0, \bar{\mathbf{r}}^{T_j} ).
\end{align*}

The function approximator consists of two stages and the same structure is used for both the stationary and maneuvering target case. In the case of a stationary target, successful `interception' means that the pursuer can reach the stationary target. The steps to evaluate
\(\tilde{f}( \mathbf{r}^{P_i}_0, \mathbf{v}^{P_i}_0, \bar{\mathbf{r}}^{T_j}_0,  \dot{\bar{\mathbf{r}}}^{T_j}_0)\) are:
\begin{enumerate}
    \item Classification of whether the pursuer can intercept the target. If not, a predetermined large time value that is expected to be larger than anything seen in practice is returned.
    \item If the pursuer can feasibly intercept the target, a regression model is used to predict the minimum time until intercept.
\end{enumerate}



Since bottleneck assignment depends only on ordering of values in the cost matrix and not on the values themselves \cite{burkard_ch6}, the predetermined large value being larger than any values output by the regression model ensures that a pursuer is not assigned to a target it cannot intercept unless no pursuer can intercept that particular target. The dependence only on ordering also means that correct classification is highly important to producing an assignment with the same bottleneck time as the truth assignment.

We implement both the classification and regression models as fully connected feed-forward neural networks. The sigmoid output of the classifier is rounded, with a 0 representing infeasible and 1 representing feasible.

\subsection{Stationary Target Network Architecture}
The classifier for a stationary target is composed of six layers. The first four layers are of size 32, each with the hyperbolic tangent activation function. These are followed by a dropout layer with probability of dropout of 0.4 to improve generalization, and finally a layer with a single output and sigmoid activation function.

The regression model for a stationary target is composed of two fully connected layers of size 128 with rectified linear unit activation functions, followed by a layer of size one with a linear activation function.

\subsection{Maneuvering Target Network Architecture}
The classifier for a maneuvering target is composed of three layers of size 128, 1024, and 128 with the hyperbolic tangent activation function, followed by a dropout layer with probability of dropout of 0.2, and a single output using the sigmoid activation function.

The regression model for a maneuvering target is composed of three layers of size 64, 1024, and 64 with the rectified linear unit activation function, followed by a layer of size one with a linear activation function.

\subsection{Model Training}

A dataset for the stationary target problem is generated by grid sampling 375,000 samples from the \(x\text{-}z\) plane with 
\(r_x \in [-15000, -5000]\ \text{ft}\), \(r_z \in [0, 30,000]\ \text{ft}\), \(v_x \in [2500, 3500]\ \text{ft/s}\), \(v_z = 0 \text{ft/s}\), \(r_{tc,x} \in [5000, 15000]\ \text{ft}\), and \(r_{tc,z} \in [0, 30,000]\ \text{ft}\). The dataset is split with \(20\%\) being taken as a test dataset, and the remaining being used for training. The dimension to be sampled over is reduced by one by translating the pursuer and target pair along the \(x\) axis such that the target is at \(r_{tc,x}=0\). The vehicle dynamics are not translation-invariant along the \(z\) axis due to change in atmospheric density. Additionally, to increase the representation of infeasible samples in the dataset, random feasible samples are removed until the fraction of feasible samples in the dataset drops from the initial \(84.6\%\) to \(80\%\). This value of \(80\%\) is found experimentally by cross-validation of a candidate set of values.

For the maneuvering target problem, we generate a training dataset by grid sampling 7,290,000 samples over the higher dimensional space where the parameters \(r_x, r_z, v_x, v_z, r_{tc,x}, r_{tc,z}\) fall in the same intervals as in the stationary target case, but we have two additional maneuvering target initial velocity parameters, \(v_{tc,x} \in [-2500, 3500]\ \text{ft/s}\) and \(v_{tc,z} = 0\ \text{ft/s}\). We reject samples where the maneuvering target is unable to reach the asset. The fraction of feasible interceptions in this dataset is 55\% so we do not remove samples to increase the proportion of feasible samples. Again, 20\% is used as a test dataset.

The asset is placed at \(\mathbf{r} = (0, 0, 5000)\ \text{ft}\). We set parameters \(u_{max} = 25\text{G}\) and \(r_{capture} = 20\text{ft}\). The optimal control problem is solved iteratively using sequential convex programming, CVX and Mosek for each convex program. 

All neural networks are implemented using Keras
and Tensorflow.
For both the stationary target and maneuvering target datasets, the classifiers are trained with a binary cross-entropy loss function and the regression models are trained with a mean squared error loss function on the subsets of the training datasets where the pursuers are able to intercept the target. All training is done with the Adam optimization algorithm on a laptop with a mobile Nvidia GTX1050.

The classifier for the stationary target has an accuracy of 98.1\%. The regression model for the stationary target achieves a mean squared error of 0.000323 and a mean absolute error of 14.81 milliseconds  on the subset of the test dataset where the pursuer can intercept the target.
Training is done with batches of size 32 for 30 epochs on the classifier and 40 epochs on the regression model and takes on the order of a minute for the classification and regression models.

The classifier for the maneuvering target has an accuracy of 99.8\%. The regression model for the maneuvering target achieves a mean squared error of 0.00181 and a mean absolute error of 10.8 milliseconds on the subset of the test dataset where the pursuer can intercept the target. Training is done with batches of size 2048 for 45 epochs on the classifier and 100 epochs on the regression model and takes on the order of 30 minutes each.

\subsection{Function Approximation Based Assignment}

Using the pre-trained function approximator \(\tilde{f}\), an approximated cost matrix \(\tilde{C}\) is constructed online such that matrix entry is computed by evaluating the function approximator instead of solving the optimal control problem:
\[
\tilde{C}_{i,j} := \tilde{f}( \mathbf{r}^{P_i}_0, \mathbf{v}^{P_i}_0, \bar{\mathbf{r}}^{T_j}_0,  \dot{\bar{\mathbf{r}}}^{T_j}_0).
\]
Then the assignment produced using this approximated cost matrix is \(Z^*_{BAP}(\tilde{C})\). The assignment problem itself is the same as the one presented in Equations \eqref{eqn:assign-obj}-\eqref{eqn:assign-z-binary}.

\section{Analysis}

\begin{table*}[t]
\caption{Average Assignment Computation Time for Stationary Targets}
\centering
\begin{tabular}{c | c c c | c c c}
\hline
\textbf{Engagement}& \multicolumn{3}{c}{\textbf{Baseline Times [ms]}} & \multicolumn{3}{c}{\textbf{Function Approximation Times [ms]}} \\
\cline{2-7}
\textbf{Size} &  Cost Matrix & Assignment & Total & Cost Matrix & Assignment & Total \\
\hline
\(3  \times  3\) & 69,300   & 0.121 & 69,300.121  & 0.369 & 0.121 & 0.490\\
\(5  \times  5\) & 161,000  & 0.182 & 161,000.182 & 1.02 & 0.182 & 1.202 \\
\(10 \times 10\) & 577,000  & 0.349 & 577,000.349 & 4.10 & 0.349 & 4.449 \\
\hline
\end{tabular}
\label{tb:assignment-computation-time-stationary}
\end{table*}

\begin{table*}[t]
\caption{Average Assignment Computation Time for Maneuvering Targets}
\begin{center}
\begin{tabular}{c | c c c | c c c}
\hline
\textbf{Engagement}& \multicolumn{3}{c}{\textbf{Baseline Times [ms]}} & \multicolumn{3}{c}{\textbf{Function Approximation Times [ms]}} \\
\cline{2-7}
\textbf{Size} &  Cost Matrix & Assignment & Total & Cost Matrix & Assignment & Total \\
\hline
\(3  \times  3\) & 69,300   & 0.121 & 69,300.121  & 0.302 & 0.121 & 0.423 \\
\(5  \times  5\) & 161,000  & 0.182 & 161,000.182 & 0.840 & 0.182 & 1.022 \\
\(10 \times 10\) & 577,000  & 0.349 & 577,000.349 & 3.360 & 0.349 & 3.709 \\
\hline
\end{tabular}
\label{tb:assignment-computation-time-maneuvering}
\end{center}
\end{table*}

We evaluate the accuracy of the function approximators by comparing assignments produced from the approximated cost matrices to assignments produced from the true cost matrices computed using the optimal control problem presented in Section \ref{sec:asset-guarding-engagements}. We consider the minimum intercept times computed using the optimal control problem to be the truth intercept times and the assignments computed from the true cost matrix to be the truth assignments. All computations are carried out on a laptop with a 2.8 GHz Intel Core i7 64-bit processor with four physical cores and 16GB of RAM.

Test sets of 500 engagements each for engagements of size 3 $\times$ 3 (three pursuers versus three targets), 5 $\times$ 5, and 10 $\times$ 10 are produced by selecting pursuer and stationary target initial conditions uniformly at random from the same region that the function approximator training dataset is sampled from. This is repeated to create three additional test sets for the maneuvering targets giving a total of six sets of 500 engagements.

We generate two cost matrices for each engagement: one, \(C\), using trajectory optimization; and the other, \(\tilde{C}\), using the function approximator. The optimization problems are solved in parallel, four at a time (one on each core). The function approximators are evaluated serially on each of the \(n^2\) pursuer-target pairs. 
We then compute the bottleneck assignments \(Z_{BAP}^*(C)\) using the true cost matrix, and \(Z^*_{BAP}(\tilde{C})\) using the approximated cost matrix.
The bottleneck problem is solved using a thresholding algorithm as in \cite{burkard_ch6}.

For both stationary and maneuvering target engagements, we compare the times needed to construct the cost matrix and solve the assignment problem for the baseline trajectory optimization method to the times needed by the function approximation method.

Table \ref{tb:assignment-computation-time-stationary} breaks down the times for the stationary target engagements. Table \ref{tb:assignment-computation-time-maneuvering} does the same for engagements with maneuvering targets. The time required to construct the cost matrix decreases by several orders of magnitude when using the function approximator, enabling the cost matrices to be computed in real-time. 
For problems of this size, the time to solve the bottleneck assignment problem is small compared to the time required to solve the optimal control problem presented in Section \ref{sec:traj-opt}.

An engagement is considered to be feasible if there exists a truth assignment such that all targets are intercepted.
The classifiers for both stationary targets and maneuvering targets are sufficiently accurate that for all feasible engagements, the assignments produced using the approximated cost matrix intercept all targets. Table \ref{tb:bottleneck-match} summarizes the fraction of feasible engagements on which the approximated cost matrix produces assignments which have the same bottleneck time as the truth bottleneck time. In these engagements, the function approximator method results in an assignment as good as the truth assignment. Note that the bottleneck assignment problem does not in general have a unique minimizer, so the approximation-based assignment may differ from the true assignment while having the same bottleneck time.

\begin{table}[b]
\caption{Fraction of Feasible Engagements with Same Bottleneck Time}
\begin{center}
\begin{tabular}{c c c}
\hline
\textbf{Engagement}&\multicolumn{2}{c}{\textbf{Percent (\%)}} \\
\cline{2-3} 
\textbf{Size} & Stationary & Maneuvering \\
\hline
\(3  \times  3\) & 98.9 & 98.5 \\
\(5  \times  5\) & 98.0 & 99.3 \\
\(10 \times 10\) & 97.0 & 95.9 \\
\hline
\multicolumn{3}{l}{}
\end{tabular}
\label{tb:bottleneck-match}
\end{center}
\end{table}

For those feasible engagements where the function approximators produce feasible assignments but with bottleneck times different than the truth bottleneck times, we measure the error by averaging the ratio of bottleneck times of the approximation-based assignments to the true bottleneck times. For a particular problem size and for either stationary or maneuvering targets, let \(Z^*_{BAP}(C_i)\) and \(Z^*_{BAP}(\tilde{C}_i)\) be the truth and model assignments for engagement \(i\), respectively. Let \(b\) be the  vectors of bottleneck times corresponding to the \(Z^*_{BAP}(C_i)\) and \(\tilde{b}\) be the bottleneck times corresponding to the \(Z^*_{BAP}(\tilde{C}_i)\):
\begin{align*}
    b_i & = J_{\text{bottleneck}}(C_i, Z^*_{BAP}(C_i)) \\  
    \tilde{b}_i & = J_{\text{bottleneck}}(C_i, Z^*_{BAP}(\tilde{C}_i)).
\end{align*}
Note that both \(b_i\) and \(\tilde{b}_i\)  are computed using the truth minimum intercept times, \(C_i\). Let \(\mathcal{I}\) be the set of indices \(i\) such that engagement \(i\) is feasible and \(b_i \neq \tilde{b}_i\), and let \(|\mathcal{I}|\) be the size of \(\mathcal{I}\). We define the bottleneck ratio for engagement \(i\) as \(\frac{\tilde{b}_i}{b_i}\). Then, we measure the mean bottleneck ratio as:
\begin{equation}
\begin{matrix}
\mathrm{mean\ bottleneck} \\ \mathrm{ratio} 
\end{matrix} \
= \
\frac{1}{|\mathcal{I}|}\sum_{i \in \mathcal{I}} \frac{\tilde{b}_i}{b_i}.
\end{equation}
In general \(\frac{\tilde{b}_i}{b_i} \geq 1\) and values closer to 1 indicate smaller error between the approximation-based assignments' bottleneck times and the truth bottleneck times. Given the restriction on \(\mathcal{I}\) that \(\tilde{b}_i \neq b_i\), we have that \(\frac{\tilde{b}_i}{b_i} > 1\) and therefore the mean bottleneck ratio will be strictly greater than one. Mean bottleneck ratios by engagement size and stationary versus maneuvering targets are summarized in Table \ref{tb:bottleneck-error}. 

\begin{table}[b]
\caption{Mean Bottleneck Ratio}
\begin{center}
\begin{tabular}{c c c}
\hline
\textbf{Engagement}&\multicolumn{2}{c}{\textbf{Mean Bottleneck Ratio}} \\
\cline{2-3} 
\textbf{Size} & Stationary & Maneuvering \\
\hline
\(3  \times  3\) & 1.3398 & 1.0036 \\
\(5  \times  5\) & 1.2040 & 1.0011 \\
\(10 \times 10\) & 1.0743 & 1.0006 \\
\hline
\multicolumn{3}{l}{}
\end{tabular}
\label{tb:bottleneck-error}
\end{center}
\end{table}



\subsection{Robustness Against Different Trajectories}

The function approximators implicitly assume that the maneuvering targets follow proportional navigation guidance with PN gain of 3. In this section we consider engagements where the maneuvering targets are following proportional navigation guidance but with PN gains of 4 and 5. PN gains of 3-5 are typical values \cite{PNratio}.

On the same test set of 500 engagements used for maneuvering engagements with PN gain of 3, we compute for each engagement two new cost matrices using trajectory optimization. One, \(C_4\), is computed simulating maneuvering target trajectories using a PN gain of 4, and the other, \(C_5\), with a gain of 5. Then, we compute the corresponding assignments \(Z^*_{BAP}(C_4)\) and \(Z^*_{BAP}(C_5)\). The function approximator does not know the PN gain used by the targets, so the function approximation-based assignment is \(Z^*_{BAP}(\tilde{C})\) regardless of true PN gain.

We compare the function approximation-based assignments to the truth assignments in the case of PN gain of 4 by computing the fraction of engagements on which the function approximation method achieves the same bottleneck time, and by computing the mean bottleneck ratio for those engagements where the function approximation method results in a different bottleneck. We repeat this for the case where the truth assignments are with PN gain of 5. The values are compared to the baseline case of the maneuvering targets following proportional navigation with a gain of 3. In all engagements where there exists a trajectory optimization-based assignment that intercepts all targets assuming that the PN gain is known, the function approximator method results in an assignment which intercepts all targets. While the fraction of engagements where the function approximation-based bottleneck exactly matches the truth bottleneck drops compared to the baseline, the mean bottleneck ratio remains close to 1. This suggests that despite the function approximators being trained on datasets where the maneuvering targets use PN gains of 3, they are robust to variance in the PN gain. Table \ref{tb:bottleneck-match-robust} summarizes the fractions of engagements for which the function approximator results in the same bottleneck time as the truth assignment by true PN gain and engagement size. Table \ref{tb:bottleneck-ratio-robust} summarizes the mean bottleneck ratios for engagements where the approximation-based bottleneck time differs from the true bottleneck time by true PN gain and engagement size.

\begin{table}[b]
\caption{Fraction of Feasible Engagements with Same Bottleneck Time vs True PN Gain}
\begin{center}
\begin{tabular}{c c c c}
\hline
\textbf{Engagement}&\multicolumn{3}{c}{\textbf{Percent (\%)}} \\
\cline{2-4} 
\textbf{Size} & Baseline (PN 3) & PN 4 & PN 5 \\
\hline
\(3  \times  3\) & 98.5 & 96.2 & 93.4\\
\(5  \times  5\) & 99.3 & 81.9 & 74.4\\
\(10 \times 10\) & 95.9 & 70.3 & 58.7\\
\hline
\multicolumn{4}{l}{}
\end{tabular}
\label{tb:bottleneck-match-robust}
\end{center}
\end{table}

\begin{table}[b]
\caption{Mean Bottleneck Ratio vs True PN Gain}
\begin{center}
\begin{tabular}{c c c c}
\hline
\textbf{Engagement}&\multicolumn{3}{c}{\textbf{Mean Bottleneck Ratio}} \\
\cline{2-4} 
\textbf{Size} & Baseline (PN 3) & PN 4 & PN 5 \\
\hline
\(3  \times  3\) & 1.0036 & 1.0305 & 1.0153 \\
\(5  \times  5\) & 1.0011 & 1.0139 & 1.0159 \\
\(10 \times 10\) & 1.0006 & 1.0113 & 1.0191 \\
\hline
\multicolumn{4}{l}{}
\end{tabular}
\label{tb:bottleneck-ratio-robust}
\end{center}
\end{table}

\section{CONCLUSIONS}

This letter presents an assignment problem scheme for assigning pursuers to stationary or maneuvering targets in a missile allocation problem, and a neural-network based function approximator that allows solving the assignment problem with reasonable accuracy in real time. This reduces the number of pursuer control trajectories to be computed from \(n^2\) to \(n\), since only the assigned trajectories need to be computed instead of all possible pairs. Effectiveness of the function approximators is demonstrated on a simulated set of engagements of various sizes.

Future work may investigate more realistic vehicle models, maneuvering targets that actively evade pursuers, unbalanced assignment problems (i.e., \(m\) vs. \(n\) engagments), engagements involving multiple assets, and extensions to dynamic task assignment.

\addtolength{\textheight}{0cm} 









\bibliography{references}{}
\bibliographystyle{ieeetr}

\end{document}